\documentstyle[preprint,aps,eqsecnum,psfig]{revtex}
\begin{document}
\tightenlines
\preprint{hep-th/0109133}
\title{ Gauss-Bonnet Black Holes in AdS Spaces}

\author{Rong-Gen Cai\footnote{Email address: cairg@itp.ac.cn}}

\address{Institute of Theoretical Physics, Chinese Academy of Sciences,
   \\
 P.O. Box 2735, Beijing 100080, China \\
Department of Physics, Osaka University, Toyonaka, Osaka 560-0043, Japan}

\maketitle
\begin{abstract}
We study thermodynamic properties and phase structures of topological
black holes in  Einstein theory with a Gauss-Bonnet term and a negative 
cosmological constant. The event horizon of these topological black holes 
can be a hypersurface with positive, zero or negative constant curvature. 
When the horizon is a zero curvature hypersurface, the thermodynamic
properties of black holes  are  completely the same as those of black holes
without the Gauss-Bonnet term, although the two black hole solutions are quite
different. When the horizon is a negative constant curvature hypersurface, the
thermodynamic properties of the Gauss-Bonnet black holes are qualitatively 
similar to those of black holes without the Gauss-Bonnet term. When the event 
horizon is a hypersurface with  positive constant curvature, we find that the 
thermodynamic properties and phase structures of black holes drastically depend 
on the spacetime dimension $d$ and the coefficient of the Gauss-Bonnet term: 
when $d\ge 6$, the properties of black hole are also qualitatively similar to
the case without the Gauss-Bonnet term, but when $d=5$, a new phase of locally
stable small black hole occurs under a critical value of the Gauss-Bonnet 
coefficient, and beyond the critical value, the black holes are always 
thermodynamically stable. However, the locally stable small black hole is not
globally preferred, instead a thermal anti-de Sitter space is globally 
preferred. We find that there is a minimal horizon radius, below which the 
Hawking-Page phase transition will not occur since for these black holes the 
thermal anti de Sitter space is always globally preferred.     
\end{abstract}

\newpage

\section{Introduction}

In recent years black holes in anti-de Sitter (AdS) spaces have attracted 
a great deal of attention. There are at least two reasons responsible 
for this.   First, in the spirit of AdS/CFT 
correspondence~\cite{Mald,Gubs,Witten1}, it has been convincingly argued by
Witten~\cite{Witten2} that the thermodynamics of black holes in AdS 
spaces (AdS black holes) can be identified with that of a certain  dual 
CFT in the high temperature limit.  With this correspondence, one can gain some
insights into thermodynamic properties and phase structures of strong 
't Hooft coupling CFTs by studying thermodynamics of AdS black holes.

Second, although the ``topological censorship theorem"~\cite{Fried} still holds 
in asymptotically AdS spaces~\cite{Wool}, it has been found that except
for the spherically symmetric black holes whose event horizon is a sphere
surface, black holes also exist with even horizon being a zero or negative 
constant curvature hypersurface. These black holes are referred to as 
topological black holes in the literature. Due to the different horizon 
structures, these black holes behave in many aspects quite different from 
the spherically symmetric black holes~\cite{Lemos}-~\cite{Aros}.  

It is by now known that the AdS Schwarzschild black hole  
 is thermodynamically unstable when the horizon radius is small,
 while it is stable for large radius; there is a phase transition, named 
Hawking-Page phase transition~\cite{Hawk}, between the large stable black hole 
and a thermal AdS space. This phase transition is explained by 
Witten~\cite{Witten2} as the confinement/deconfinement transition of  
the Yang-Mills theory in the AdS/CFT correspondence.  However, 
it is interesting to note that if event horizon of AdS black holes is a 
hypersurface with zero or negative constant curvature, the black hole
is always stable and the corresponding CFT is always dominated by the
black hole. That is, there does not exist the Hawking-Page phase transition
for AdS black holes with a Ricci flat or hyperbolic horizon~\cite{Birm}.

Higher derivative curvature terms occur in many occasions, such as in  the
semiclassically quantum gravity and in the effective low-energy
action of superstring theories. In the latter case, according to the
AdS/CFT correspondence, these terms can be viewed as the corrections of
large $N$  expansion of boundary CFTs in the strong coupling limit.
Due to the non-linearity of Einstein equations, however, it is very difficult
to find out nontrivial exact analytical solutions of the Einstein equations 
with these higher derivative terms. In most cases, one has to adopt some 
approximation methods  or find solutions numerically.

Among the gravity theories with higher derivative curvature terms, the so-called 
Lovelock gravity~\cite{Love} is of
some special features in some sense. For example, the resulting field 
equations  contain no more than second derivatives of the metric and it has
been proven to be free of ghosts when expanding about the flat space, 
evading any problems with unitarity. The Lagrangian of Lovelock theory is
the sum of dimensionally extended Euler densities\footnote{The gravity theory with
a Gauss-Bonnet term was originally proposed by Lanczos~\cite{Lanc}, independently
rediscovered by Lovelock~\cite{Love}. See also discussions in \cite{Zumi} and 
\cite{Zwie}.}
\begin{equation}
\label{1eq1}
{\cal L} =\sum^n_ic_i{\cal L}_i,
\end{equation}
where $c_i$ is an  arbitrary constant and ${\cal L}_i $ is the Euler density
of a $2i$-dimensional manifold,
\begin{equation}
\label{1eq2}
{\cal L}_i=2^{-i}\delta^{a_1b_1\cdots a_ib_i}_{c_1d_1\cdots c_id_i}
  R^{c_1d_1}_{\ \ a_1b_1}\cdots R^{c_id_i}_{\ \ a_ib_i}.
\end{equation}
Here ${\cal L}_0=1$ and hence $c_0$ is just  the cosmological    
constant. ${\cal L}_1$ gives us the usual Einstein-Hilbert term and ${\cal L}_2$
is the Gauss-Bonnet term.  A spherically symmetric static solution 
of (\ref{1eq1}) has been found in the sense that the metric function is 
determined by solving for the real roots of a polynomial 
equation~\cite{Whee}. Since the Lagrangian (\ref{1eq1}) includes
many arbitrary coefficients $c_i$,  it is difficult to extract
physical information from the solution. In Refs.~\cite{Banados2,Zanelli}, by 
restricting these coefficients to a special set so that the metric
function can be readily determined by solving the polynomial equation, 
some exact, spherically symmetric black hole solutions have been found. 
Black hole solutions with nontrivial topology in this theory have been also 
studied in Refs.~\cite{Cai2,Aros}.

In this paper we will analyze black hole
solutions in  Einstein theory with a Gauss-Bonnet term and a negative
cosmological constant, in which the Gauss-Bonnet coefficient is not fixed.
In this theory a static, spherically symmetric black hole solution was first
discovered by Boulware and Deser~\cite{Deser}. However, the thermodynamic properties
of the solution were not discussed there. Here we will first generalize this 
solution to the case in which the horizon of black holes can be 
a positive, negative or zero constant curvature hypersurface, and then discuss
thermodynamic properties and phase structures of black holes.
Because of this Gauss-Bonnet term, some nontrivial and interesting features 
will occur.

\section{Topological Gauss-Bonnet black holes}

The Einstein-Hilbert action with a Gauss-Bonnet term and a 
negative cosmological constant, $\Lambda=-(d-1)(d-2)/2l^2$, 
in $d$ dimensions can be written down 
as\cite{Deser}~\footnote{The Gauss-Bonnet term is a 
topological invariant in four dimensions. So $d\ge 5$ is assumed 
in this paper.}  
\begin{equation}
\label{3eq1}
S=\frac{1}{16\pi G}\int d^dx\sqrt{-g}\left(R +\frac{(d-1)(d-2)}{l^2}
  + \alpha (R_{\mu\nu\gamma\delta}R^{\mu\nu\gamma\delta} -4 R_{\mu\nu}
   R^{\mu\nu}+R^2)\right),
\end{equation}
where $\alpha$ is the Gauss-Bonnet coefficient with dimension $(length)^2$ 
and is positive in the heterotic string theory~\cite{Deser}. So we restrict
ourselves to the case $\alpha \ge 0$~\footnote{We will make a simple discussion
for the case $\alpha <0$ in Sec.~III.}. Varying the action yields the 
equations of gravitational field
\begin{eqnarray}
\label{3eq2}
R_{\mu\nu}-\frac{1}{2}g_{\mu\nu}R &= &\frac{(d-1)(d-2)}{2l^2}g_{\mu\nu}
  + \alpha \left (\frac{1}{2}g_{\mu\nu}(R_{\gamma\delta\lambda\sigma}
  R^{\gamma\delta\lambda\sigma}-4 R_{\gamma\delta}R^{\gamma\delta}
  +R^2) \right. \nonumber \\
 &&~~~- \left. 2 RR_{\mu\nu}+4 R_{\mu\gamma}R^{\gamma}_{\ \nu}
  +4 R_{\gamma\delta}R^{\gamma\  \delta}_{\ \mu\ \ \nu}
   -2R_{\mu\gamma\delta\lambda}R_{\nu}^{\ \gamma\delta\lambda} \right).
\end{eqnarray}
We assume the metric being of the following
form
\begin{equation}
\label{3eq3}
ds^2 = -e^{2\nu}dt^2 +e^{2\lambda}dr^2 +r^2 h_{ij}dx^idx^j,
\end{equation}
where $\nu$ and $\lambda$ are functions of $r$ only,  
and $h_{ij}dx^idx^j$ represents the line element of a 
($d-2$)-dimensional hypersurface with constant curvature 
$(d-2)(d-3)k$ and volume $\Sigma_k$.
 Without loss of the generality, one may take $k=1$, $-1$ or $0$.
Following Ref.~\cite{Deser} and substituting the ansatz (\ref{3eq3}) into 
the action (\ref{3eq1}), we obtain
\begin{equation}
\label{3eq4}
S=\frac{(d-2)\Sigma_k}{16\pi G}\int dt\ dr e^{\nu+\lambda}\left [ r^{d-1}
 \varphi (1+\tilde \alpha \varphi) +\frac{r^{d-1}}{l^2}\right]',
\end{equation}
where a prime denotes derivative with respect to $r$, 
$\tilde\alpha =\alpha (d-3)(d-4)$ and $\varphi =r^{-2}(k- e^{-2\lambda})$.   
From the action (\ref{3eq4}), one can find the solution
\begin{eqnarray}
&& e^{\nu +\lambda}=1, \nonumber \\
&& \varphi (1+\tilde\alpha \varphi)+\frac{1}{l^2} = 
   \frac{16\pi G M}{(d-2)\Sigma_k r^{d-1}}, 
\end{eqnarray}
from which we obtain the exact solution\footnote{It is not so obvious 
that the minisuperspace approach applies for non-spherically symmetric 
solutions in the gravity theory. However, it can be checked that the solution
(\ref{3eq6}) indeed satisfies the equations (\ref{3eq2}) of motion. This is related
to the fact that following \cite{Wilt1}, one can show that a Birkhoff-like theorem 
holds in the gravity theory (\ref{3eq1}).}
\begin{equation}
\label{3eq6}
e^{2\nu} =e^{-2\lambda}=k +\frac{r^2}{2\tilde\alpha}\left ( 1 \mp
 \sqrt{1+\frac{64 \pi G\tilde \alpha M}{(d-2)\Sigma_k r^{d-1}}-
   \frac{4\tilde\alpha}{l^2}} \right), 
\end{equation}
where $M$ is the gravitational mass of the solution\footnote{This gravitational mass 
can be obtained by substituting the solution (\ref{3eq6}) into the action (\ref{3eq4}) 
and then using boundary term method. For this method, for example, see \cite{Zanelli}.}.
 The solution with $k=1$ and
spherical symmetry was first found by Boulware and Deser~\cite{Deser}.
Here we extend this solution to include the cases $k=0$ and $-1$.
Note that the solution (\ref{3eq6}) has two branches with ``$-$" or
``$+$" sign. Moreover, there is a potential singularity at the place where
the square root vanishes in (\ref{3eq6}), except for the singularity 
at $r=0$. Here we mention that the theory (\ref{3eq1}) with $\tilde\alpha =l^2/4$
discussed in Ref.~\cite{Zanelli} in five dimensions  corresponds to the 
theory proposed in \cite{Cham1}; the solution with $\tilde\alpha=l^2/4$ 
discussed in Refs.~\cite{Banados2,Zanelli} in five dimensions was also 
included in Refs.~\cite{Cai2,Aros}. 
 If $\tilde\alpha=0$, namely,
without the Gauss-Bonnet term, the solution (\ref{3eq6}) reduces to the one in
\cite{Birm}, and the thermodynamics of the latter was discussed there.

When $M=0$, the vacuum solution in (\ref{3eq6}) is
\begin{equation}
\label{3eq7}
e^{-2\lambda}= k+\frac{r^2}{2\tilde\alpha}\left(1 \mp \sqrt{1-
   \frac{4\tilde\alpha}{l^2}}\right).
\end{equation}
Since $\tilde\alpha >0$, one can see from the above that $\tilde \alpha$ must
obey $4\tilde\alpha /l^2 \le 1$, beyond which this theory is undefined. Thus,
 the action (\ref{3eq1}) has two 
AdS solutions with effective cosmological constants $l^2_{\rm eff}=
\frac{l^2}{2}\left( 1 \pm \sqrt{1-\frac{4\tilde \alpha}{l^2}}\right)$. 
When $4\tilde\alpha/l^2=1$, these two solutions coincide with each other, 
resulting in $e^{-2\lambda}= k+2r^2/l^2$ and that the theory has a unique 
AdS vacuum~\cite{Banados2,Zanelli}. On the other hand, if 
$\tilde\alpha<0$, the solution (\ref{3eq7}) is still an AdS space if one
takes the "$-$" sign, but becomes a de Sitter space if one takes 
the  "$+$" sign and $k=1$. From the vacuum case, the solution (\ref{3eq7})
with both signs seems reasonable, from which we cannot determine which sign
in (\ref{3eq6}) should be adopted.  This problem can be solved by considering
the propagation of gravitons on the background (\ref{3eq7}). It has been
shown by Boulware and Deser~\cite{Deser} that the branch with ``$+"$ sign is
unstable and the graviton is a ghost, while the branch with ``$-$" 
sign is stable and is free of ghosts. This can also be seen from the 
case $M \ne 0$.  When $k=1$ and $1/l^2=0$, just as observed by Boulware and 
Deser~\cite{Deser}, the solution is asymptotically a Schwarzschild solution 
if one takes the ``$-$" sign, but is asymptotically an AdS  Schwarzschild 
solution with a negative gravitational mass for the ``$+$'' sign, indicating the
instability. Therefore the branch with ``$+$" sign in (\ref{3eq6}) is of less 
physical interest\footnote{A detailed analysis of the solution (\ref{3eq6})
without the negative cosmological constant, namely, $1/l^2=0$, has been made 
in \cite{Myers,Wilt}.}. From now on, we will not consider the branch 
with ``$+$" sign.

From (\ref{3eq6}), the mass of black holes can be expressed in terms of
the horizon radius $r_+$,
\begin{equation}
\label{3eq8}
M=\frac{(d-2)\Sigma_k r_+^{d-3}}{16\pi G}\left (k +\frac{\tilde\alpha k^2}
  {r_+^2} +\frac{r_+^2}{l^2}\right).
\end{equation}
The Hawking temperature of the black holes can be easily obtained by requiring
the absence of conical singularity at the horizon in the Euclidean sector 
of the black hole solution. It is
\begin{equation}
\label{3eq9}
T =\left. \frac{1}{4\pi}\left(e^{-2\lambda}\right)'\right |_{r=r_+} 
 = \frac{(d-1)r_+^4 +(d-3)kl^2 r_+^2 +(d-5)\tilde \alpha k^2l^2}
   {4\pi l^2 r_+(r_+^2+2\tilde\alpha k)}.
\end{equation}
Usually entropy of black holes satisfies the so-called area formula. This is,
the black hole entropy equals to one-quarter of horizon area. 
In gravity theories with higher derivative curvature terms, however, 
in general the entropy of black
holes does not satisfy the area formula. To get the black hole entropy, 
in \cite{Cai2} we suggested a simple method according to the fact that as a 
thermodynamic system, the entropy of black hole must obey the first law of
black hole thermodynamics: $dM = TdS$.
Integrating the first law, we have 
\begin{equation}
\label{3eq10}
S = \int T^{-1}dM =\int ^{r_+}_0 T^{-1}\left (\frac{\partial M}
     {\partial r_+}\right) dr_+,
\end{equation}
where we have imposed the physical assumption that the entropy vanishes when
the horizon of black holes shrinks to zero\footnote{Note that for the $k=-1$ black hole,
there exists a minimal horizon radius. For these black holes, therefore the horizon cannot 
shrink to zero. However, it is known that the black hole entropy is a function of horizon
surface~\cite{Wald}. According to the second law of black hole mechanics, the black hole 
entropy can be expressed in terms of a polynomial of horizon radius $r_+$ with positive 
exponents. As a result, although the black hole horizon cannot shrink to zero when $k=-1$,
this method seems applicable as well. The results in \cite{Cai2} and in this paper show 
this point. For example, when $\tilde \alpha =0$, the formula (\ref{3eq11}) gives the entropy
of AdS black holes in Einstein theory without the Gauss-Bonnet term. Obviously, in this
case the resulting area formula (\ref{3eq11}) holds as well in the case of 
$k=-1$.}. Thus once given the temperature
and mass of black holes in terms of the horizon radius, One can readily get the
entropy of black holes and needs not know in which gravitational theory the 
black hole solutions are. Substituting (\ref{3eq8}) and (\ref{3eq9}) into 
(\ref{3eq10}), we find the entropy of the Gauss-Bonnet black holes
 (\ref{3eq6}) is 
\begin{equation}
\label{3eq11}
S = \frac{\Sigma_k r_+^{d-2}}{4 G}\left( 1+\frac{(d-2)}{(d-4)}
    \frac{2 \tilde \alpha k}{r_+^2}\right).
\end{equation}
When $k=1$, it is in complete agreement with the one in \cite{Myers}, there
the entropy of the Gauss-Bonnet black holes without the cosmological constant
is obtained by calculating the Euclidean action of black holes. 
The heat capacity of black holes is
\begin{equation}
\label{3eq12}
C=\left(\frac{\partial M}{\partial T}\right) =\left(
    \frac{\partial M}{\partial r_+}
  \right) \left(\frac{\partial r_+}{\partial T}\right),
\end{equation}
where
\begin{eqnarray}
\label{3eq13}
 \frac{\partial M}{\partial r_+} &=&
   \frac{(d-2) \Sigma_k}{4G}r_+^{d-5}(r_+^2 +2\tilde\alpha k)T, \nonumber \\
\frac{\partial T}{\partial r_+}
 &=& \frac{1}{4\pi l^2 r_+^2(r_+^2+2\tilde\alpha k)^2}
     \left [ (d-1)r_+^6 -(d-3)kl^2 r_+^4 +6(d-1)k\tilde \alpha r_+^4
     \right. \nonumber \\
&& +2(d-3)\tilde\alpha k^2l^2 r_+^2 
      -3 (d-5) \tilde\alpha k l^2 r_+^2 -2(d-5) \tilde \alpha^2 k^2 l^2 ].
\end{eqnarray}
The free energy of black holes, defined as $F=M-TS$, is
\begin{eqnarray}
\label{3eq14}
F &=& \frac{\Sigma_k r_+^{d-5}}{16\pi G (d-4)l^2 (r_+^2 +2\tilde \alpha k)}
   \left [-(d-4)r_+^6 +(d-4)kl^2 r_+^4  \right. \nonumber \\
  &&~~~~ -6(d-2)k \tilde \alpha r_+^4 +(d-8)\tilde \alpha k^2l^2 r_+^2
  +2(d-2)\tilde \alpha^2 k l^2].
\end{eqnarray}
Thus we give  some thermodynamic quantities of Gauss-Bonnet 
black holes in AdS spaces. When $\tilde \alpha =0$, these thermodynamic quantities 
reduce to corresponding ones in Ref.~\cite{Birm}. In Fig.~1 the inverse temperature
$\beta =1/T$ of the black holes versus the horizon radius is plotted. We can see
clearly different behaviors for the cases $k=1$, $0$ and $-1$: The inverse 
temperature always starts from infinity and monotonically decreases to zero in the 
cases $k=0$ and $k=-1$, while it starts from zero  and reaches its maximum at a certain 
horizon radius and then goes to zero monotonically when $k=1$. This indicates that
for the cases $k=-1$ and $k=0$, the black holes are not only locally thermodynamic 
stable, but also globally preferred, while in the case of $k=1$, the black hole 
is not locally thermodynamic stable for small radius, but it is for large radius.
Therefore, for the $k=1$ case, there is a Hawking-Page phase transition. For details
see \cite{Birm}.

When $\tilde\alpha \ne 0$,  we see that those quantities drastically
depend on the parameter $\tilde\alpha$, horizon structure $k$ and the
spacetime dimension $d$. Below we will discuss each case according to
the classification of horizon structures, $k=0$, $k=-1$ and $k=1$, 
respectively.

\subsection{The case of $k=0$}

In this case we have
\begin{eqnarray}
\label{3eq15}
&& T= \frac{(d-1) r_+}{4\pi l^2}, \nonumber \\
&& S= \frac{\Sigma_k}{4G} r_+^{d-2}, \nonumber \\
&& C= \frac{(d-2)\Sigma_k}{4G} r_+^{d-2}, \nonumber \\
&& F= -\frac{\Sigma_k}{16\pi G} \frac{r_+^{d-1}}{l^2},
\end{eqnarray}
where $r_+^{d-1}=16\pi  G l^2 M/(d-2)\Sigma_k$. It is interesting to note that 
these thermodynamic quantities are independent of the parameter $\tilde\alpha$. That 
is, these quantities have the completely same expressions as 
those \cite{Birm} for black holes without the Gauss-Bonnet term. We therefore 
conclude that in the case 
$k=0$, the black holes with and without Gauss-Bonnet term have completely 
same thermodynamic properties, although the two solutions are quite different,
which can be seen from (\ref{3eq6}). In particular, we note
here that the entropy of the Gauss-Bonnet black holes still satisfies the
area formula in the case $k=0$.

\subsection{The case of $k=-1$}

As the case \cite{Birm} without the Gauss-Bonnet term, there are also so-called
``massless" black hole and ``negative" mass black hole in the  Gauss-Bonnet 
black hole (\ref{3eq6}). When $M=0$, the black hole has the horizon radius
\begin{equation}
\label{3eq16}
r_+^2 =\frac{l^2}{2}\left( 1\pm \sqrt{1-\frac{4\tilde \alpha}{l^2}}\right), 
\end{equation}
with Hawking temperature $T=1/2\pi r_+$. Here there are two ``massless" black 
hole solutions, corresponding to two branches in the solution (\ref{3eq6}). 
But
the black hole with smaller horizon radius belongs to the unstable branch.

Given a fixed $\tilde\alpha$, the smallest black hole has the horizon 
radius 
\begin{equation}
\label{3eq17}
r^2_{\rm min}
   =\frac{(d-3)l^2}{2(d-1)}\left( 1+\sqrt{1-\frac{(d-1)(d-5)}{(d-3)^2}
   \frac{4\tilde\alpha}{l^2}}\right).
\end{equation}
The black hole is an extremal one, it has vanishing Hawking temperature  and
the most ``negative" mass
\begin{equation}
\label{3eq18}
M_{\rm ext}= -\frac{(d-2)(d-3)\Sigma_k l^2 r_{\rm min}^{d-5}}{16\pi G (d-1)^2}
  \left ( 1-\frac{d-1}{d-3} \frac{4\tilde\alpha}{l^2}
  +\sqrt{1-\frac{(d-1)(d-5)}{(d-3)^2}\frac{4\tilde\alpha}{l^2}}\right).
\end{equation}
When $4\tilde \alpha/l^2 =1$, the smallest radius is $r^2_{\rm min}=l^2/2$ 
and $M_{\rm ext}=0$, independent of the spacetime dimension $d$. But in this
case, the Hawking temperature does not vanish. It is $T=1/\sqrt{2}\pi l$.
This is an exceptional case.

From the solution (\ref{3eq6}), one can find that in order for the solution
to have a black hole horizon, the horizon radius must obey
\begin{equation}
\label{3eq19}
r_+^2 \ge 2 \tilde \alpha.
\end{equation}
Thus the smallest radius (\ref{3eq17}) gives a constraint on the allowed
value of the parameter $\tilde\alpha$: 
\begin{equation}
\label{3eq20}
r^2_{\rm min} \ge 2\tilde \alpha,
\end{equation}
which leads to $4\tilde\alpha/l^2 \le 1$.
Since the theory is defined in the region $4 \tilde\alpha/l^2 \le 1$, the
condition (\ref{3eq20}) is always satisfied.
Due to the existence of the smallest black holes (\ref{3eq17}), we see
from (\ref{3eq9}) that except for the case $4\tilde\alpha /l^2 =1$, 
the temperature of black hole always starts from zero at the smallest radii,
corresponding to the extremal black holes and  monotonically goes to infinity
as $r_+ \to \infty$.  In the case $ 4\tilde \alpha /l^2 =1$, the temperature
starts from $1/\sqrt{2}\pi l$ at $r_+^2=l^2/2$. This can also be verified 
by looking at the behavior of the heat capacity (\ref{3eq12}). After 
considering the fact that $r_+^2 \ge 2 \tilde \alpha$ and $4\tilde\alpha/l^2 
\le 1$, it is easy to show that the heat capacity is always positive. In Fig.~2
we plot the inverse temperature of black holes in six dimensions versus the
parameter $\tilde\alpha/l^2$ and the horizon radius $r_+/l$.   

Among the smallest black holes (\ref{3eq17}), the most smallest one is 
 $r_+^2=l^2/2$ when $4\tilde \alpha/l^2=1$, its
free energy is zero. Therefore the free energy is always negative for
other black holes since the heat capacity is always positive.
As a result, the thermodynamic properties of the  black holes with 
the Gauss-Bonnet term are qualitatively similar to those of black holes
without the Gauss-Bonnet term: These black holes are always stable not
only locally, but also globally.  

In addition, let us note that except for the singularity at $r=0$,  the 
black hole solution (\ref{3eq6}) has another singularity at
\begin{equation}
\label{3eq21}
r_s^{d-1}=\frac{4\tilde \alpha r_+^{d-3}}{1-4\tilde\alpha/l^2}
    \left( 1-\frac{\tilde \alpha}{r_+^2} -\frac{r_+^2}{l^2}\right),
\end{equation}
when $M_{\rm ext} < M <0$ . But both singularities are shielded by the 
event horizon $r_+$.

\subsection{The case of $k=1$}
    
This case is very interesting. From the temperature (\ref{3eq9}) one can see
that the case $d=5$ is quite different from the other cases $d\ge 6$. When 
$d=5$, the temperature starts from zero at $r_+=0$ and goes to infinity
as $r_+\to \infty$, while it starts from infinity at $r_+=0$  as $d\ge 6$. 
In Fig.~3 we show the inverse temperatures of black holes with 
$\tilde\alpha/l^2=0.001$ in different dimensions $d=5$, $6$ and $d=10$, 
respectively. The behavior of temperature of black holes with the Gauss-Bonnet
term in $d\ge 6$ dimensions is similar to that of AdS black holes without the
Gauss-Bonnet term. But the case of $d=5$ (see Fig.~4) is quite different
from the corresponding one without the Gauss-Bonnet term (see Fig.~1). Comparing
Fig.~4 with Fig.~1, we see that a new phase of stably small black hole occurs in
the Gauss-Bonnet black holes. 

When $d=5$, we have from (\ref{3eq8}) the black hole horizon
\begin{equation}
r_+^2 =\frac{l^2}{2}\left(-1 +\sqrt{1+\frac{4(m-\tilde\alpha)}{l^2}}\right),
\end{equation}
where $m=16\pi G M/3\Sigma_k$. Therefore, in this case there is a mass gap 
$M_0=3\Sigma_k\tilde\alpha /(16\pi G)$: all black holes have a mass 
$M \ge M_0$. Using the horizon radius, from Fig.~4 we can see that the
black holes can be classified to three branches: 
\begin{eqnarray}
&& {\rm branch\ 1}:\ \   0 < r_+ < r_1, \ \ \ C>0, \nonumber \\
&& {\rm branch\ 2}:\ \  r_1 < r_+ < r_2,\ \ \ C<0, \nonumber \\
&& {\rm branch\ 3}:\ \  r_2 < r_+ < \infty,\ \ \ C>0,
\end{eqnarray}
where 
\begin{equation}
r_{1,2}^2 =\frac{l^2}{4}\left(1-\frac{12\tilde\alpha}{l^2}\right)
   \left (1 \mp \sqrt{1-\frac{16\tilde\alpha}{l^2}\left( 1-
   \frac{12\tilde\alpha}{l^2}\right)^{-2}}\right).
\end{equation}
with the assumption $36 \tilde\alpha/l^2 <1$. In the branch 1 and 3, the
heat capacity is positive, while it is negative in the branch 2. Therefore
the black holes are locally stable in the branch 1 and 3, and unstable in the
branch 2. At the joint points of branches, namely, $r_+=r_{1,2}$, the heat
capacity diverges. Comparing with the case without the Gauss-Bonnet term, 
one can see that the branch 1 is new.

When $\tilde \alpha$ increases to the value, $\tilde\alpha/l^2 =1/36$, we 
find that the branch 2 with negative heat capacity disappears. Beyond this
value, the heat capacity is always positive and the Gauss-Bonnet black holes 
are always locally stable. In Fig.~5, we show the inverse temperatures of 
Gauss-Bonnet black hole with the parameter $\tilde\alpha/l^2$, subcritical
value $0.001$, critical value $1/36$, and supercritical value $0.20$, 
respectively. 
In Fig.~6, the continuous evolution of the inverse temperature is plotted with
the parameter $\tilde\alpha/l^2$ from zero to $0.25$, from which one can see 
clearly that the black holes evolve from two branches to one branch via 
three branches.  
 
However, inspecting the free energy (\ref{3eq14}) reveals that these 
stably small black holes are not globally preferred: The free energy always
starts from some positive value at $r_+=0$ and then goes to negative infinity
as $r_+ \to \infty$. In Fig.~7 the free energy of black holes with different
parameter $\tilde\alpha/l^2$ is plotted. We see that all curves cross the 
horizontal axis (horizon radius) one time only, where $F=0$. 
In Fig.~8 we plot the region where the free energy is negative.
The region is 
\begin{equation}
\label{3eq25}
\tilde\alpha_1 < \tilde \alpha <\tilde\alpha_2,
\end{equation} 
where
\begin{equation}
\tilde\alpha _{2,1}=\frac{r_+^2}{4}+\frac{3r_+^4}{2l^2}
   \pm \frac{r_+^2}{2}\sqrt{ \frac{9r_+^4}{l^4} +\frac{11r_+^2}{3l^2}
    -\frac{5}{12}}.
\end{equation}
The joint point of the two curves is at $\tilde\alpha/l^2 =0.0360$ and 
$ r_+/l= 0.3043 $. Beyond this region, the thermal AdS space is globally
preferred. We see that there is a smallest horizon radius $r_+/l =0.3043$:
there will not exist the Hawking-Page phase transition when the black hole
horizon is smaller than the value $r_+/l =0.3043$. When black holes cross 
the curves $\tilde\alpha_2$ and $\tilde\alpha_1$, a Hawking-Page phase 
transition happens.

The region in which black holes are locally stable is determined by the 
curve $\tilde\alpha_0$,
\begin{equation}
\tilde\alpha_0 = \frac{l^2 r_+^2 -2 r_+^4}{2 l^2 +12 r_+^2}.
\end{equation}
In Fig.~9 the curve $\tilde \alpha_0$ is plotted (the lowest one): the region
is locally stable  above this curve, namely, $\tilde \alpha > \tilde \alpha_0$,
and locally unstable below this curve. 
This curve $\tilde\alpha_0$ touches the curve $\tilde \alpha_1$ at 
$\tilde\alpha=1/36\approx 0.0278$ and $r_+/l=0.4082$. Unfortunately, in 
Fig.~9 most part of the curve $\tilde \alpha_2$ is outside the plot. In
Fig.~9 one can see that there is a large region where black holes are locally
stable, but not globally preferred.

When $d\ge 6$, unlike the case $d=5$, there is no the mass gap.
 The properties of Gauss-Bonnet black holes are qualitatively
similar to those of black holes without the Gauss-Bonnet term. This can be 
seen from the behavior of the Hawking temperature of black holes in Fig.~3.
This implies that the equation $\frac{\partial T}{\partial r_+}=0$ has only 
one positive real root $r_+ = r_0(d,\tilde\alpha/l^2)$. Using (\ref{3eq13}), 
one can obtain the positive real root. But its expression is complicated, so 
we do not present it here. Given a  spacetime dimension $d$ and a 
fixed parameter $\tilde \alpha /l^2$, when a black hole has a horizon 
$r_+ >r_0$, the black hole is locally stable. Otherwise, it is unstable.  

The free energy (\ref{3eq14}) always starts from zero in the case $d\ge 6$,
reaches a positive maximum at some $r_+$, and then goes to negative infinity
as $r_+ \to \infty$. This behavior is the same as the case without the 
Gauss-Bonnet term (see the curve of $\tilde \alpha=0$ in Fig.~7). The region
where the black hole is globally preferred is restricted by a relation like 
(\ref{3eq25}), but with
\begin{eqnarray}
\tilde\alpha_{2,1} &=&\frac{r_+^2}{4(d-2)l^2}\left[ 6(d-2)r_+^2-(d-8)l^2
     \right. \nonumber \\
   &&~~~~~~~~~~\pm \sqrt{36(d-2)^2r_+^4-4(d-2)(d-16)l^2r_+^2 
  +d(32-7d)l^4}\ ].
\end{eqnarray}
And as in the case of $d=5$, these two curves connect at
\begin{eqnarray}
&& r_+^2 =\frac{l^2}{18(d-2)}\left( d-16 +\sqrt{(d-16)^2 +9d (7d-32)}
  \right), \nonumber \\
&&
\tilde\alpha =\frac{r_+^2}{4(d-2)l^2}\left( 6(d-2)r_+^2-(d-8)l^2\right).
\end{eqnarray}
in the $\tilde\alpha -r_+$ plane. Therefore the phase structure of black holes
in $d \ge 6$ dimensions is similar to the one in $d=5$ dimensions (Fig.~8).

Finally let us mention that the temperature behavior (Fig.~4) of $d=5$ 
Gauss-Bonnet black holes is quite similar to the one of the 
Reissner-Nordstr\"om (RN) black holes in AdS spaces in the canonical 
ensemble~\cite{Cham,Cvetic}. There under the critical value of charge, 
the phase of stably small black holes occurs as well. However, there is a 
big difference between two cases: For the RN black holes, the small black 
hole is not only
locally stable, but also globally preferred, while the small Gauss-Bonnet
black hole is only locally stable and  not globally preferred, instead a
thermal AdS space is preferred.

\section{Conclusions and Discussions}

We have presented exact topological black hole solutions in 
Einstein theory with a Gauss-Bonnet term and a negative cosmological
constant, generalizing the spherically symmetric black hole solution
found by Boulware and Deser~\cite{Deser} to the case where 
the event horizon of black holes is a positive, zero or negative constant 
curvature hypersurface. We have examined thermodynamic properties and analyzed 
phase structures of these topological black holes.

When the even horizon is a zero curvature hypersurface, we find that 
thermodynamic properties of Gauss-Bonnet black holes are completely the same 
as those without the Gauss-Bonnet term, although the two black hole solutions
are quite different. As a result, these $k=0$ Gauss-Bonnet black holes are 
not only locally thermodynamic stable, but also globally preferred. In 
particular, the entropy of these black holes satisfies the area formula. Note
that usually black holes in gravity theories with higher derivative curvature terms
do not obey the area formula.

When the even horizon is a negative constant hypersurface, these black holes
are qualitatively similar to those without the Gauss-Bonnet term. These 
$k=-1$ Gauss-Bonnet black holes are always locally stable and globally
preferred.

When the event horizon is a positive constant hypersurface, however, some
interesting features occurs. When $d=5$, a new phase of thermodynamically
stable small black holes appears if the Gauss-Bonnet
coefficient is under a critical value.  Beyond the critical value, the black 
holes are always thermodynamically stable. Inspecting the free energy of black 
holes  reveals these stable small black holes are not globally preferred, 
instead a thermal AdS space is preferred. The phase structures are plotted, 
from which we find that there is a smallest black hole radius. Beyond this
radius, the Hawking-Page phase transition will not happen.  From the 
phase diagram we see that there is a large region in which the black hole is
locally stable, but not globally preferred.  When $d \ge 6$, however, the 
new phase of stable small black holes disappears. Once again,  
the thermodynamic properties of the black holes are qualitatively similar
to those of black holes without the Gauss-Bonnet term.

Now we discuss the case $\alpha <0$. The vacuum solution is still the one
(\ref{3eq7}). So in this case there is no restriction on the value of 
$\tilde \alpha$; the solution (\ref{3eq6}) is still asymptotically AdS. 
Those expressions of thermodynamic quantities (\ref{3eq8}), (\ref{3eq9}),
(\ref{3eq11}), (\ref{3eq13}) and (\ref{3eq14}) are applicable as well.

When $k=0$, since thermodynamic quantities are independent of the parameter 
$\tilde \alpha$ in this case, the conclusion is the same as the
case $\tilde\alpha >0$, but with a new singularity at
\begin{equation}
r_s^{d-1}= \frac{4|\tilde\alpha|/l^2}{1+4|\tilde\alpha|/l^2}
  r_+^{d-1},
\end{equation}
which is always shielded by the event horizon $r_+$.

When $k=-1$, the
situation is similar to the case $\tilde\alpha >0$, nothing special 
appears. In this case, the smallest radius  is
\begin{equation}
r_{\rm min}^2= \frac{(d-3)}{2(d-1)}l^2 \left( 1 +\sqrt{1 +\frac{(d-5)(d-1)}
  {(d-3)^2}\frac{4|\tilde\alpha|}{l^2}}\right).
\end{equation}
The smallest black hole has a vanishing Hawking temperature.
Inside the event horizon there is an additional singularity at
\begin{equation}
r_s^{d-1}=\frac{4|\tilde\alpha|/l^2}{1+4|\tilde\alpha|/l^2}
    r_+^{d-1} \left( 1 -\frac{l^2}{r_+^2}-\frac{|\tilde\alpha|l^2}{r_+^4}
   \right),
\end{equation}
except for the one at $r=0$. The black holes are also always 
locally stable and globally preferred.

When $k=1$, there is also a smallest horizon radius
\begin{equation}
\label{min}
r^2_{\rm min} = 2|\tilde\alpha|,
\end{equation}
but this smallest black hole has a divergent Hawking temperature. In this case
the event horizon coincides with an additional singularity at 
$r^2=2|\tilde\alpha|$. For larger black holes the additional singularity is 
located at
\begin{equation}
r_s^{d-1}=\frac{4|\tilde\alpha|/l^2}{1+4|\tilde\alpha|/l^2}
    r_+^{d-1} \left( 1 +\frac{l^2}{r_+^2}-\frac{|\tilde\alpha|l^2}{r_+^4}
   \right), 
\end{equation}
inside the black hole horizon. The inverse temperature of black holes starts
from zero at the smallest radius (\ref{min}), reaches its maximal at some $r_+$
and goes to zero when $r_+ \to \infty$.  The thermodynamic properties of 
black holes are qualitatively similar to the case with $\tilde\alpha=0$.  As
a result, the new phase, which appears in the case $d=5$ and $0 <\tilde 
 \alpha/l^2 \le  1/36$, does not occur in this case.

{\bf Note added}: Perturbative AdS black hole solution with $R^2$ curvature terms and its
thermodynamic properties have been discussed recently in \cite{Noji}.

\section*{Acknowledgments}
This work was supported in part by a grant from Chinese Academy of Sciences,
and in part by the Japan Society for the
Promotion of Science and Grants-in-Aid for Scientific Research
Nos. 99020, 12640270.

\begin{figure}
\psfig{file=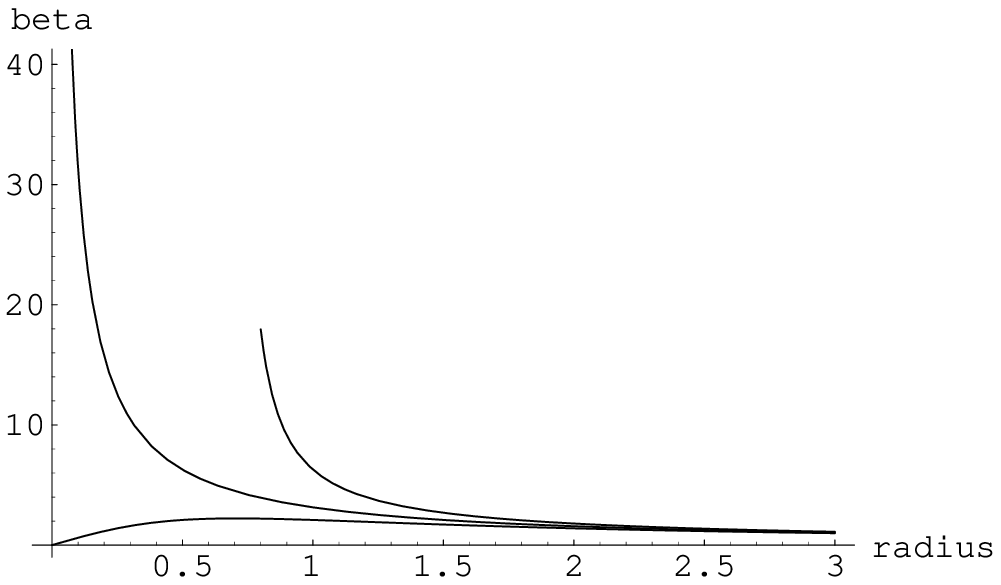,height=55mm,width=120mm,angle=0}
\caption{The inverse temperature of topological black holes without the 
Gauss-Bonnet term. The three curves above from  up to 
down  correspond to the cases $k=-1$, $0$ and $1$, respectively.}
 \end{figure}

\begin{figure}
\psfig{file=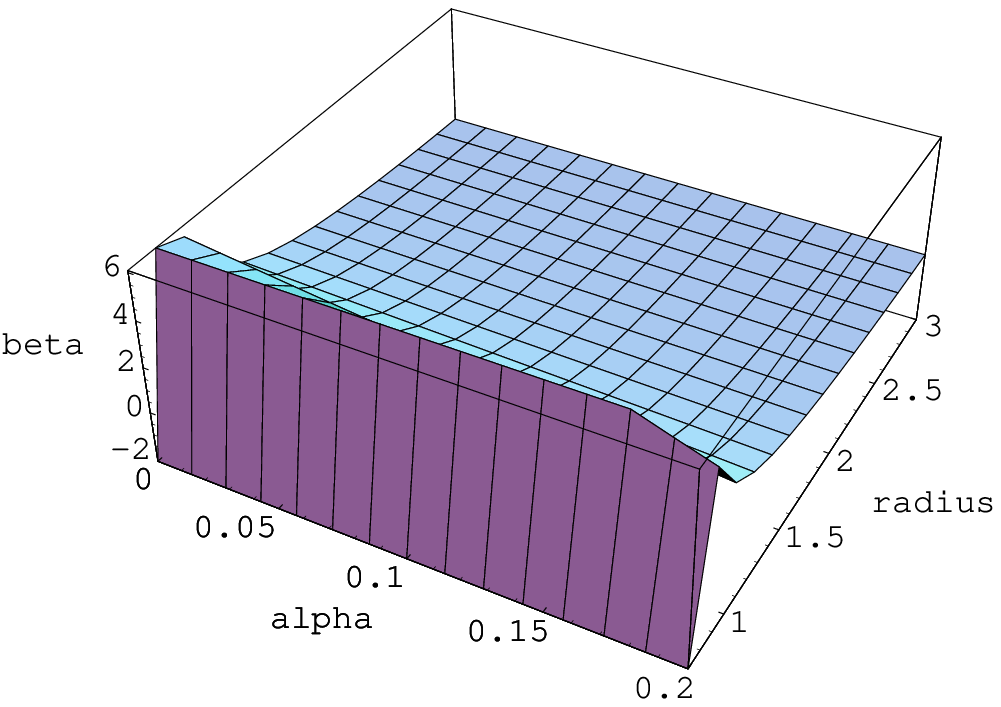,height=55mm,width=120mm,angle=0}
\caption{The inverse temperature of the $k=-1$ Gauss-Bonnet black holes in  
$d=6$ dimensions.}
\end{figure} 

\begin{figure}
\psfig{file=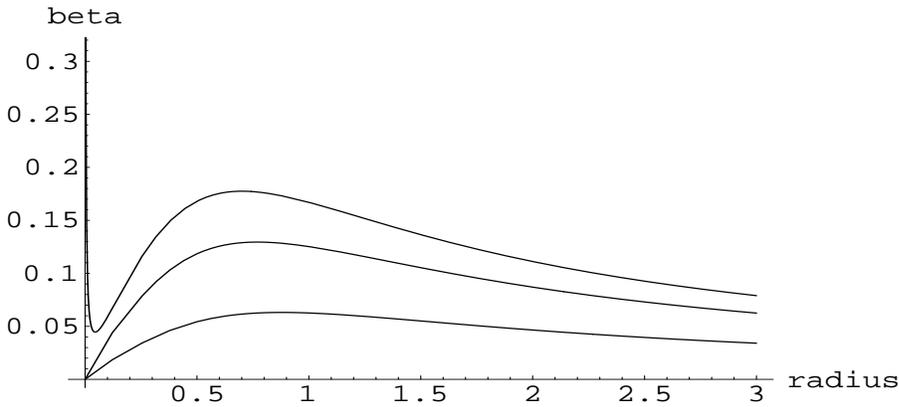,height=55mm,width=120mm,angle=0}
\caption{The inverse temperature of the $k=1$ Gauss-Bonnet black holes with 
$\tilde\alpha/l^2 =0.001$. The three curves from up to down correspond to
 $d=5$, $6$ and $d=10$, respectively. }
\end{figure} 

\begin{figure}
\psfig{file=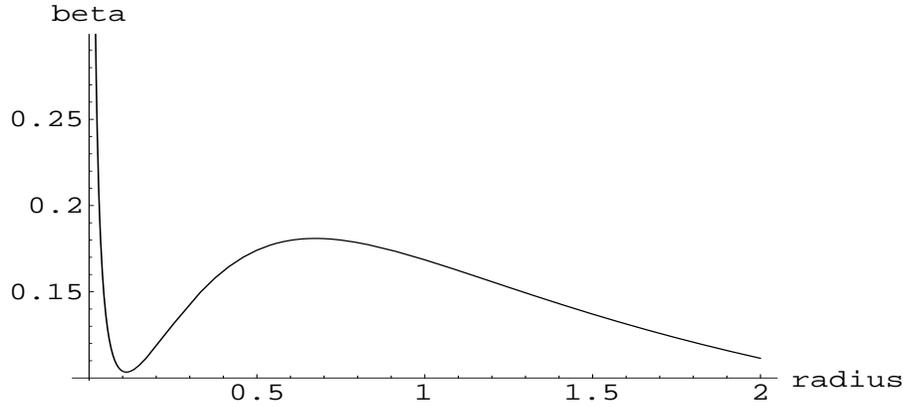,height=55mm,width=120mm,angle=0}
\caption{The inverse temperature of the $k=1$ Gauss-Bonnet black holes 
in $d=5$ dimensions with $\tilde\alpha/l^2 =0.0056$.}
\end{figure}

\begin{figure}
\psfig{file=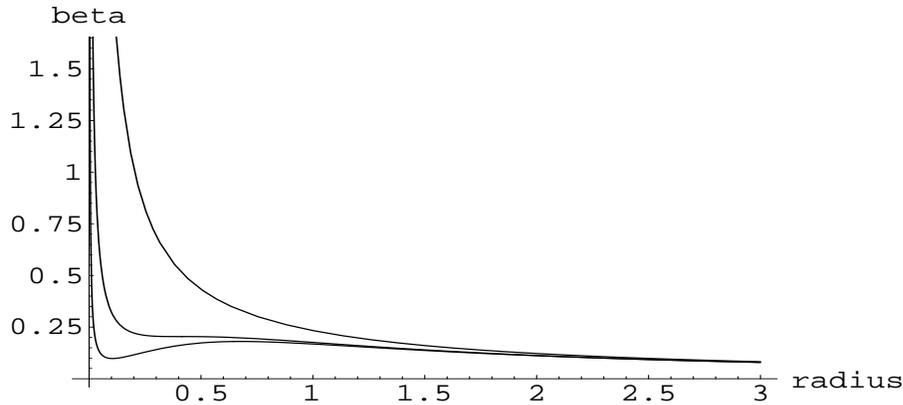,height=55mm,width=120mm,angle=0}
\caption{The inverse temperature of the $k=1$ Gauss-Bonnet black holes in
 $d=5$ dimensions. The three curves from up to down correspond to 
the cases with the supcritical $\tilde\alpha/l^2= 0.20$, critical $1/36 
\approx 0.0278 $, and subcritical $0.005$, respectively.}
\end{figure} 

\begin{figure}
\psfig{file=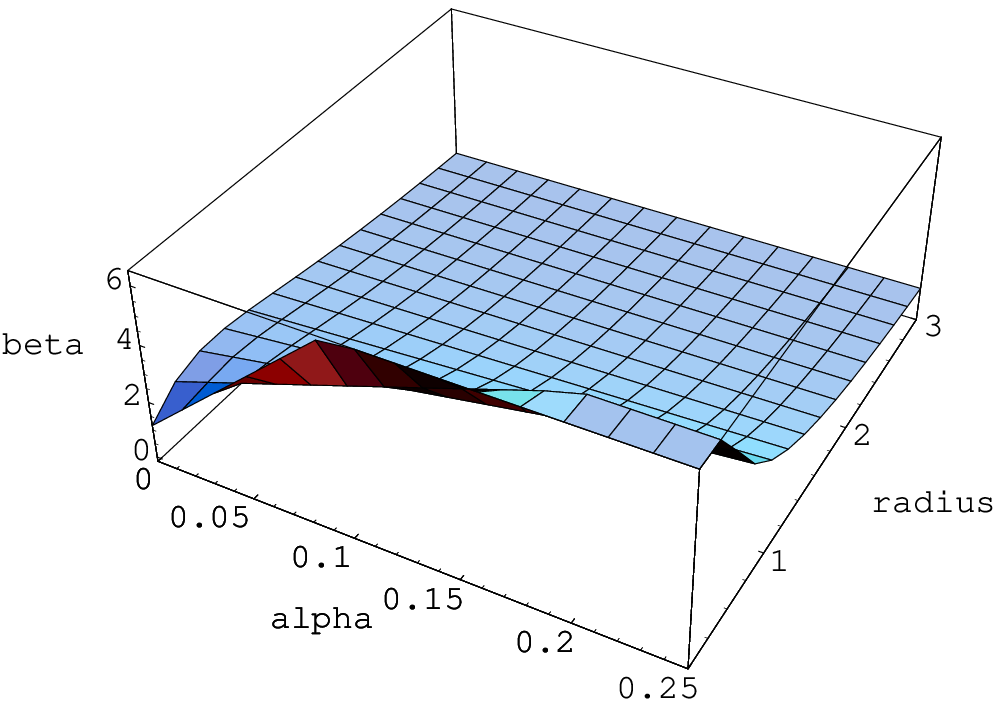,height=60mm,width=120mm,angle=0}
\caption{The inverse temperature  of the $k=1$ 
 Gauss-Bonnet black holes in $d=5$ dimensions.}
\end{figure}

\begin{figure}
\psfig{file=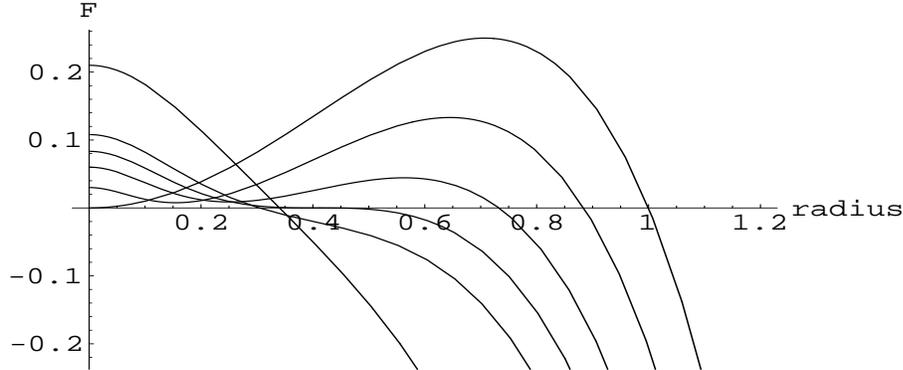,height=50mm,width=120mm,angle=0}
\caption{The free energy of the $k=1$ Gauss-Bonnet black holes in
 $d=5$ dimensions. The curves counting up to down on the F-axis 
 correspond to the cases
 $\tilde\alpha/l^2 =0.070$, $0.036$, $1/36$, $0.020$, $0.010$ and $0$, 
 respectively.}
\end{figure}

\begin{figure}
\psfig{file=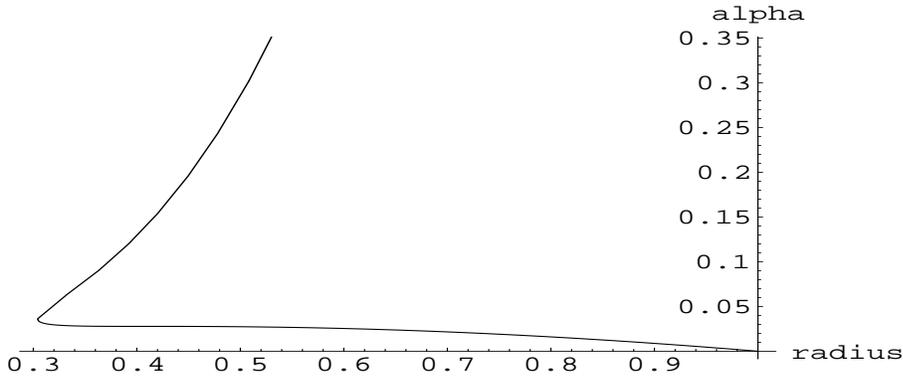,height=50mm,width=120mm,angle=0}
\caption{The curves $\tilde\alpha_2$ (the upper one) and $\tilde\alpha_1$ (the
lower one) for the Gauss-Bonnet black holes in $d=5$ diemnsions. They  
joint at $r_+/l=0.3043$ and $\tilde\alpha/l^2=0.0360$. In the region between
 $\tilde\alpha_2$ and $\tilde\alpha_1$  black holes have a negative free energy
and are globally perferred.}
\end{figure}

\begin{figure}
\psfig{file=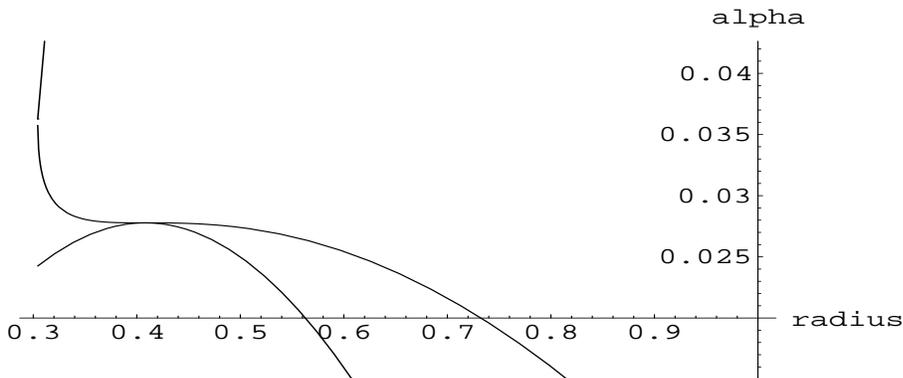,height=50mm,width=120mm,angle=0}
\caption{The curves $\tilde\alpha_2$, $\tilde\alpha_1$ and $\tilde\alpha_0$ (
the lowest one) for the Gauss-Bonnet black holes in $d=5$ dimensions. The 
region above the curve $\tilde\alpha_0$ is locally stable. The curve 
$\tilde\alpha_0$  touches the curve $\tilde\alpha_1$ at $r_+/l=0.4082$ and 
$\tilde\alpha/l^2 =0.0278$. The seperated one is the curve $\tilde\alpha_2$.}
\end{figure}

\end{document}